# Theoretical and experimental evidence of a site-selective Mott transition in Fe₂O₃ under pressure

E. Greenberg,[1,a,b] I. Leonov,[2,3,a] S. Layek,[1] Z. Konopkova,[4] M. P. Pasternak,[1] L. Dubrovinsky,[5] R. Jeanloz,[6] I. A. Abrikosov,[7,3] G. Kh. Rozenberg[1]

[1] *School of Physics and Astronomy, Tel Aviv University, 69978, Tel Aviv, Israel.*
[2] *Theoretical Physics III, Center for Electronic Correlations and Magnetism, Institute of Physics, University of Augsburg, 86135 Augsburg, Germany*
[3] *Materials Modeling and Development Laboratory, NUST "MISIS," 119049 Moscow, Russia*
[4] *DESY, HASYLAB, PETRA-III, P02, Notkestr. 85, Bldg. 47c, Hamburg, Germany.*
[5] *Bayerisches Geoinstitut, University of Bayreuth, Bayreuth, Germany.*
[6] *Department of Earth and Planetary Science, University of California, Berkeley, California 94720*
[7] *Department of Physics, Chemistry and Biology (IFM), Linköping University, SE-581 83 Linköping, Sweden*

[a] E.G. and I.L. equally contributed to this work
[b] E.G.'s current affiliation is GSEACRS, University of Chicago, Argonne, IL, USA.

We provide experimental and theoretical evidence for a novel type of pressure-induced insulator-metal transition characterized by *site-selective* delocalization of the electrons. Mössbauer spectroscopy, X-ray diffraction and electrical transport measurements on Fe₂O₃ up to 100 GPa, along with density functional plus dynamical mean-field theory (DFT+DMFT) calculations, reveal this site-selective Mott transition between 50 and 68 GPa, such that the metallization can be described by $(^{VI}Fe^{3+})_2O_3$ [ $R\bar{3}c$ structure]

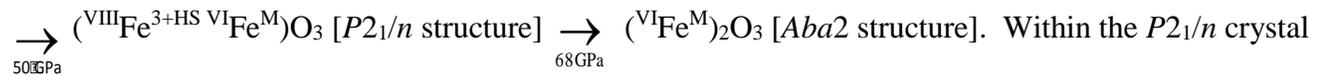

$(^{VIII}Fe^{3+HS}~^{VI}Fe^{M})O_3$ [$P2_1/n$ structure] $\xrightarrow{68GPa}$ $(^{VI}Fe^{M})_2O_3$ [$Aba2$ structure]. Within the $P2_1/n$ crystal structure, characterized by two distinct coordination sites (VI and VIII), we observe equal abundances of ferric ions (Fe³⁺) and ions having delocalized electrons (Fe^M), and only at higher pressures is a fully metallic *Aba*2 structure obtained, all at room temperature. Thereby the transition is characterized by delocalization/metallization of the *3d* electrons on half the Fe sites, with a site-dependent collapse of local moments. Above ~50 GPa, Fe₂O₃ is a strongly correlated metal with reduced electron mobility (large band renormalizations) of $m^*/m \sim 4$ and 6 near the Fermi level. Importantly, upon decompression, we observe a site-selective (metallic) to conventional Mott insulator phase transition

$(^{VIII}Fe^{3+HS}~^{VI}Fe^{M})O_3$

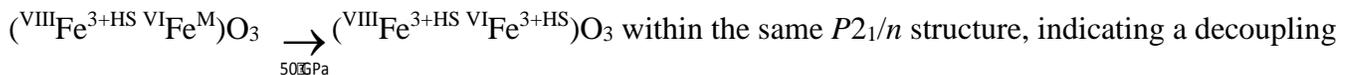

$(^{VIII}Fe^{3+HS}~^{VI}Fe^{3+HS})O_3$ within the same $P2_1/n$ structure, indicating a decoupling of the electronic and lattice degrees of freedom, characteristic of a true Mott transition. Our results provide a new paradigm for understanding insulator-metal transitions in correlated electron materials, and show that the interplay of electronic correlations and lattice may result in rather complex behavior of the electronic structure and magnetic state of such compounds.

The insulator-metal transition, induced by pressure, composition or by other means, represents perhaps the most profound transformation of the chemical bond in materials. A specific subset, the Mott transition, is of particular importance because it is thought to depend on electron correlations that are essential to understanding the properties of transition-metal oxides important to fields ranging from materials chemistry to condensed-matter physics and even planetary science. Electronic and magnetic transitions in strongly correlated transition-metal compounds have thus been among the main topics of condensed-matter research over recent decades, being especially relevant to understanding high-temperature superconductivity as well as heavy-fermion behavior.[1,2,3,4,5,6,7] The definitive electronic phenomenon in such compounds is the breakdown of $d$- or $f$-electron localization, causing a Mott (Mott-Hubbard) insulator-to-metal transition [1,2]. Such a transition does not necessarily imply a rearrangement of atoms, but in fact often exhibits an appreciable collapse in volume[8,9,10,11]. The initial concept of Mott is based on the relative importance of kinetic hopping (measured by the bandwidth) and onsite repulsion of electrons. However, recently another mechanism was proposed suggesting that a change of the crystal-field splitting rather than variation in the bandwidth may drive a Mott transition[10,12,13,14]. As a result, the Mott transition involves a simultaneous insulator–metal transition, magnetic moment collapse (change of the local spin state) and volume collapse.

Along with this, the existence in real materials of many additional degrees of freedom may result in new scenarios for the Mott transition; e.g., orbital degrees of freedom give rise to a scenario of an orbital-selective Mott transition[15,16,17,18]. Nevertheless, in some materials these paradigms cannot explain the experimentally observed details of electronic and structural transformations. An outstanding example is $Fe_2O_3$ hematite (Néel temperature $T_N$ =956 K [19]), which crystallizes in a corundum-type structure with one type of $FeO_6$ octahedron (slightly distorted). Photoemission spectroscopy measurements show that $Fe_2O_3$ exhibits Mott insulating behavior with a large energy gap of ~ 2.5 eV[20]. Upon compression, $Fe_2O_3$ is known to undergo a first-order phase transition above ~50 GPa, which is accompanied by a remarkable collapse of the lattice volume by ~10 %[,21,22,23,24,25,26,]. The phase transition was associated with a "partial" transition to low-spin (LS) $Fe^{3+}$ state[24,27,28,29,30]. However, later Mössbauer studies indicated that at $P > 80$ GPa only a non-magnetic phase exists, without any sign of magnetic moments down to the lowest temperatures achieved, thereby suggesting a breakdown of *3d-electron* localization[25]. Furthermore, electrical transport measurements[25] show a sharp insulator-metal transition at a substantially lower pressure of ~50 GPa, which seemingly contradicts the existence of the high-spin state at pressures far above this transition (the high-spin state would not be expected to exhibit the electrical conductivity of a metal)[24,25,27,28].

The crystal structure of the high-pressure phase of $Fe_2O_3$ observed above ~50 GPa has been assigned either to an orthorhombic perovskite[22,24,31] or a $Rh_2O_3$-II-type[23,25] structure in early studies. Only recently, based on single-crystal diffraction studies[32,33], the lattice structure has been defined as a double-perovskite (DPv) phase. It has been proposed[23,34] that this structural transition drives the electronic and magnetic transformation. On the other hand, recent theoretical DFT+DMFT calculations[35] (DFT+DMFT: density-functional theory + dynamical mean-field theory of correlated electrons[36]) predict that the electronic transition occurs within the hematite phase, i.e. prior to the structural transition, at high compression of $V < 0.8\ V_0$ ($V_0$ is the equilibrium unit cell volume). According to an experimental equation of state (EOS), e.g.[22,23], such volumes correspond to $P > 70$ GPa. We note that in ref. [35] the structural complexity of $Fe_2O_3$ near the phase transition has not been considered. Therefore the interplay between correlated electrons and the lattice structure could not have been studied. Thus, despite significant efforts and dozens of publications on this outwardly simple system, the evolution with pressure of the electronic and crystallographic structure of $Fe_2O_3$, and the mechanism of its electronic transition remain unresolved. Current theoretical models[35] do not explain the existing experimental results[24, 25,27,28,29,30,34].

In the present work, we have studied $Fe_2O_3$ to pressures of about 100 GPa, combining Mössbauer spectroscopy, electrical transport and X-ray diffraction upon compression and decompression, along with first-principles quantum mechanical calculations (DFT+DMFT). Our study reveals a new type of insulator-metal transition, providing what is to the best of our knowledge the first evidence for a *site-selective* Mott transition characterized by delocalization and, hence, metallization of the Fe *3d* electrons on only half of the Fe sites within the crystallographic unit cell: a transition accompanied by collapse of the local moments on the same octahedral Fe sites of the distorted DPv structure. This behavior clearly distinguishes the transformation in $Fe_2O_3$ from conventional and orbital-selective Mott metal-insulator transitions. The DPv phase of $Fe_2O_3$ is a strongly correlated metal with reduced mobility (high effective mass, *m\**) of electrons near the Fermi energy, *m\**/*m* ~ 4 to 6, and upon decompression it reverts back to a conventional Mott insulator (*m* is the normal mass of the electron). Our work highlights the interplay between electronic and crystalline structures, and not only explains a long-standing controversy regarding the high-pressure behavior of $Fe_2O_3$ but also suggests that *site-selective* Mott transitions can be typical of transition-metal compounds.

## Results

### Mössbauer spectroscopy

Mössbauer spectra of $Fe_2O_3$, characteristic of various pressure ranges and recorded at room and low temperatures, are shown in Figure 1(a, b, c). In agreement with previous publications[24,25,27,28], the only observed spectral component upon compression is that of the high-spin state up to nearly 48 GPa (hematite phase Figure 1(a)). At $P \geq 48$ GPa two new *equally abundant* components emerge: a non-magnetic quadrupole-split component, with no sign of magnetic correlations down to 8 K (Figure 1(b)), and a magnetically split component characterized by significantly reduced hyperfine field $H_{hf}$ (SI Figure S1). Taking into account the reported double perovskite-type structure in this pressure range[32], the two components are designated by $DPv_{nm}$ and $DPv_m$, respectively. At $P \geq 56$ GPa the only spectral components are DPv$_{nm}$ and DPv$_m$ with equal abundances, until at $P \geq 62$ GPa the abundance of a non-magnetic component starts to increase (Figure 2(a)) to the point that above 75 GPa the Mössbauer spectra show a single, quadrupole-split component, designated as the high-pressure (HP) state (Fig. 1(c)). The lack of any signs of magnetic correlations on Mössbauer timescales ($\approx 10^{-7}$ s) down to 4 K (Figure 1(c)) prompted us to designate this single HP component as a non-magnetic state.

Upon decompression, the DPv components reappear, with a hysteresis of ~6 GPa (Figure 2(b)). In the pressure range of $47 > P > 44$ GPa, the two DPv components are replaced by a single HS component designated as DPv$_{dec}$ (Figs. 1(d), S1). However, a component with hyperfine parameters identical to those of the hematite phase appears only below ~35 GPa, with complete transition back to hematite at ~25 GPa (Figs. 1(d), 2).

**X-ray diffraction**

Powder and single-crystal X-ray diffraction were performed at room temperature, respectively, during compression to 62 GPa followed by decompression, and up to 101 GPa on compression alone[37]. Our studies show that upon compression a first-order structural phase transition with a symmetry change takes place in the pressure range of 53 to 57 GPa, with a concomitant volume decrease of ~9% (Figures 3, S2, S3). Single-crystal diffraction[32,33] shows that the new intermediate-pressure phase can be described as a distorted double perovskite-type (DPv- Fe$_2$O$_3$) using a monoclinic unit-cell with $P2_1/n$ symmetry[32] (the symmetry is actually triclinic $P$-$I$[33], however we use a monoclinic model to constrain the atomic arrangement as done in ref. 33). This structure has the general formula $A_2B'B''O_6$ and consists of a three-dimensional network of tilted corner-sharing $B'$O$_6$ and $B''$O$_6$ octahedra with $A$-cations located in bicapped trigonal-prismatic voids (8 nearest neighbors) (Figure S4, interatomic distances are given in SI Table 1). The unit-cell volume as a function of pressure for hematite and DPv- Fe$_2$O$_3$ is shown in Figure 3, combining both powder and single-crystal[33] data. It is noteworthy that *upon decompression* from 62 GPa the DPv phase remains stable down to 35 GPa (Figures 3, SI Figures S2, S3). However, at $P \sim 50$ GPa a discontinuous volume increase of ~7% is observed, while the DPv structure remains unchanged (Figure 3) and only the monoclinic distortion of the unit cell decreases (SI Figure S5). The structural transition back to hematite begins only below 35 GPa.

A transition to a new high-pressure polymorph is observed upon compression of the DPv-$Fe_2O_3$ phase above 67 GPa (Figure 3), with a diffraction pattern that is successfully indexed based on an orthorhombic $Aba$2 space group that has only one type of Fe cation[33] (SI Figure S4).

**Electrical resistance measurements**

Our electrical resistance measurements show an abrupt ~6 orders of magnitude decrease of resistance at about 40-60 GPa on compression (Fig. 4), in agreement with the reported insulator-metal transition at ~55 GPa [25,38]. Upon further compression, we observe a substantial change in the pressure dependence of the resistance, indicating an additional change of conductivity features at ~70 GPa (Fig. 4(a)). Similar behavior is seen during the decompression cycle, with a hysteresis of ~ 6 GPa. It is noteworthy that upon pressure release the resistance rises only by ~3 orders of magnitude at about 50-40 GPa, saturating below 40 GPa. Furthermore, to avoid a structural transition back to the corundum structure, we terminated decompression at 37 GPa and performed recompression measurements up to 83 GPa. The pressure-temperature dependence of electrical resistance upon recompression shows an abrupt drop at 45-60 GPa, with the onset of metallization below 53 GPa (Fig. 4(b)). Similar to the hematite phase[38] the temperature dependence of the resistance of the insulating DPv phase is associated with a variable-range hopping mechanism below ~50 GPa: the electrical conductivity varies as $\sigma = C$ $\exp(T_0/T)^{1/4}$, though we notice a significantly reduced Mott temperature value, $T_0$, (Fig. 4(b); SI Fig. S6(b)). Meanwhile, within the metallic region above ~50 GPa, the resistance exhibits a clear deviation from the Fermi-liquid-like $T^2$ behavior, showing a minimum at temperature $T = 110 - 150$ K for the DPv phase and at about 75 K for the HP phase (Fig. S6(a), for details see Supporting Info). This behavior is presumably associated with a Kondo effect[39].

**DFT+DMFT calculations**

We further combined our experimental results with computations of the electronic structure and spin state of $Fe_2O_3$ by employing the DFT+DMFT approach, a combination of *ab initio* band structure methods and dynamical mean-field theory of correlated electron systems[36]. We employ a fully charge self-consistent implementation of the DFT+DMFT method and compute the electronic structure and phase stability of $Fe_2O_3$ under pressure. As a starting point, we calculate the electronic and structural properties of the low-pressure *R-3c* phase in the paramagnetic state (i.e., at slightly elevated temperature, $T = 390$ K, that should not otherwise affect our conclusions). In agreement with previous studies[35], we obtain a Mott insulating solution with a relatively large *d-d* energy gap of about 2.5 eV. The calculated equilibrium lattice constant 5.61 a.u. and bulk modulus ~187 GPa are in good agreement with the X-ray diffraction measurements. Our result for the local magnetic moments is ~4.76 $\mu_B$, documenting that at ambient pressure the $Fe^{3+}$ ions are in a high-spin state ($S = 5/2$) with localized *3d* electrons. We note that our calculations predict the HS-LS transition in the *R-3c* $Fe_2O_3$ to occur upon compression above ~72 GPa, i.e., at substantially higher compression than ~50 GPa found in our experiments (see Supplementary Material p.10).

Upon compression above ~50 GPa, the corundum *R-3c* phase undergoes a structural transition to the DPv phase. We calculate the electronic properties of that phase using the monoclinic *P2₁/n* symmetry and crystal structure parameters as obtained by diffraction at ~54 GPa, and use a cluster expansion of the DFT+DMFT approach in order to treat correlations in the Fe *3d* bands of the structurally distinct Fe *A* and Fe *B* sites. Our results for the spectral function (Fig. 5) reveal the existence of a *site-selective* Mott phase, in which the *3d* electrons on only half of the Fe sites (octahedral *B* sites) are metallic, while the *A* sites remain insulating. The possibility of a site-selective Mott phase has recently been discussed for the rare-earth nickelates, but without experimental or theoretical confirmation[40,41]. We note that the Fe *B* $a_{1g}$ and $e_g^\pi$ orbitals show a sharp quasiparticle peak at the Fermi level, which is associated with a pronounced (orbital-selective) enhancement of the effective electron mass, *m\*/m*. In fact, we estimate *m\*/m* ~ 6 for the Fe $B$ $a_{1g}$ and ~ 4 for the $e_g^\pi$ orbitals at temperature ~390 K. In contrast to that, the Fe $B$ $e_g^\sigma$ orbitals remain insulating. In agreement with our experiment, we observe that the insulator-to-metal transition is accompanied by a remarkable site-selective collapse of local moments. Indeed, the local magnetic moment at the Fe $A$ sites is ~4.63 $\mu_B$, which differs substantially from the magnetic moment at the $B$ sites ~ 0.89 $\mu_B$. This confirms that in the DPv phase the octahedral Fe $B$ ions are in a low-spin state ($S = 1/2$), while the Fe $A$ sites remain high-spin ($S = 5/2$). Moreover, our results for the spin-spin correlation function $\chi(\tau)$ = $\langle m_z(\tau)m_z(0)\rangle$ (see inset of Figure 5) show that the $3d$ electrons on the Fe $A$ ions are localized to form fluctuating moments ($\chi(\tau)$ is seen to be almost constant and close to unity). In contrast, the $3d$ electrons on the Fe $B$ ions exhibit a rather itinerant magnetic behavior for the $a_{1g}$ and $e_g^\pi$ orbitals, implying a localized to site-selective itinerant moment transition in $Fe_2O_3$ under pressure. We note that this remarkable combination of localized and itinerant $3d$ electrons can give rise to a complex electronic state of DPv $Fe_2O_3$ at low temperatures, e.g., resulting in a heavy Fermion-like behavior associated with the Kondo effect. Indeed, our experimental transport data exhibit a Kondo-like abnormal behavior of resistance at ~ 110-150 K, while our calculations predict a substantial enhancement of the effective mass of $3d$ electrons and show a sharp Kondo-like peak at the Fermi level. Furthermore, this behavior is consistent with the observed absence of magnetic correlations of the Fe B sites in the MS spectra down to the lowest measured temperatures. The Fe $B$ electrons are delocalized, the corresponding long-time magnetic susceptibilities are well screened and the (instantaneous) amplitude of the fluctuating moments is small. Because of that, no magnetic response can be detected by a relatively slow probe such as Mössbauer spectroscopy.

We note that upon lattice expansion of the DPv structure by ~8.5 %, our calculations predict a phase transition into a *conventional* Mott insulating state. Indeed, our Mössbauer, electrical resistance and X-ray diffraction measurements show that upon decompression the electronic and structural transitions do not coincide; they are separated by a pressure interval of ~20 GPa. This implies a transition between site-selective and conventional Mott phases for the DPv structure upon decompression and recompression: a remarkable decoupling of the electronic and lattice degrees of freedom in $Fe_2O_3$. We conclude that we are documenting intrinsic electronic properties of the DPv phase of $Fe_2O_3$, as indicated by the small hysteresis in electrical resistance upon decompression and recompression near the onset of insulator-metal transition (Fig. 4: contrast the recompression results for hematite in previous studies[25,38]).

**Discussion**

Summarizing our theoretical and experimental results, we find evidence that the metallization transition in $Fe_2O_3$ occurs in stages with pressure, first for half the Fe cations in the DPv phase - those in the octahedral $B'$ and $B''$ sites with the collapsed magnetic moment (DPvnm), while the prismatic Fe $A$ sites remain insulating and high-spin, and then for all the Fe in the high-pressure $Aba2$ structure. Assigning iron ions contributing electrons to the conduction band as a nominal $Fe^M$, we can summarize the transitions as ($^{VI}Fe^{3+HS}$)$_2O_3$ [$R\bar{3}c$ ] $\rightarrow$ ($^{VIII}Fe^{3+HS\ VI}Fe^M$)$O_3$ [$P2_1/n$] $\rightarrow$ ($^{VI}Fe^M$)$_2O_3$ [$Aba2$] with increasing pressure, where subscript Roman numerals indicate nearest-neighbor (Fe–O) coordination and crystal structures are given in brackets (HS designates high spin for the ferric ion). The average Fe–O bond length collapses upon metallization, from 1.91 Å in hematite at 51 GPa to 1.82 Å in the high-pressure $Aba2$ phase at 74 GPa; in between, it is the coexistence of small octahedral sites along with large 8-coordinated sites in the DPv structure ($P2_1/n$, with average bond lengths of 1.78-1.86 Å and 2.09 Å, respectively, at 54 GPa) that allows for the site-selective insulator-metal transition (SI Table 1). The volume change upon metallization is identical to that observed in $CaFe_2O_4$[42], suggesting a similar mechanism of electronic transition for these sites: namely, closure of the Mott-Hubbard gap associated with a spin transition[10,12,13] in accord with our theoretical calculations. Because the DPv structure of $Fe_2O_3$ consists of chains of octahedra linked along the crystallographic $c$ direction, and separated by the 8-coordinate sites in the $a$ and $b$ directions (SI Fig. 4), we would expect this phase to exhibit anisotropic electron mobility, with higher conductivity in the $c$ than the $a$ or $b$ directions.

We note that metallization does not occur in the hematite phase upon compression. In the region where the MS data find both the hematite and the DPv phases, the remaining hematite phase is still in a high-spin state (even when half of the Fe in the DPv phase are already non-magnetic, metallic). In addition, the $P(V)$ data in Fig 3 show that there is no appreciable change in unit-cell volume of hematite during compression (no deviation from the hematite EOS), even in the region of coexistence. This is in full accordance with our theoretical calculations, which show that in hematite the IMT associated with a HS-LS state transformation takes place at pressure $\sim$ 72 GPa (at volume $\sim$ 0.74 $V_0$) (SI Fig. 7).

Upon decompression, we observe a sharp reversal in electronic properties at about 45 GPa, with a metal-to-insulator transition and retrieval of a magnetic state (Figures 1c and 4), which can be described by: $(^{VIII}Fe^{3+HS}\ ^{VI}Fe^{M})O_3\ [P2_1/n] \rightarrow (^{VIII}Fe^{3+HS}\ ^{VI}Fe^{3+HS}\ [P2_1/n])O_3$. Though, the DPv structure remains unchanged and only the volume increases by $\sim$7%, documenting the decoupling of electronic and structural phase transitions. Thus, our decompression study provides, to the best of our knowledge, the first observation of a "true" site-selective Mott transition, not complicated by any coinciding structural transitions. The structural transition from DPv back to the original corundum-type structure takes place only below 35 GPa (Figures 3, S2).

Formation of an intermediate electronic state, the so-called "orbital-selective" phase, has been claimed to occur in multi-orbital transition-metal oxides[15,16,17,18]. In the orbital-selective phase, due to the inclusion of orbital degrees of freedom, a partial localization can take place, in which some orbitals are conducting, while others are localized. As a result, localized spins and itinerant electrons coexist in a system. In contrast to materials with orbital-selective state(s), in $Fe_2O_3$ the reason for the appearance of an intermediate electronic state, with a coexistence of localized (HS) and itinerant (non-magnetic) Fe *3d* electrons, is the occurrence of a site-selective Mott phase made possible by the two, distinct (6-fold and 8-fold) coordination environments.

Indeed, the appearance of the $P2_1/n$ DPv phase can be understood as a result of the interplay between cohesive (lattice) energy and local magnetic moments. While the former favors the denser high-pressure phase (e.g., *Aba2*), the local magnetic moments enter into the total energy as $\propto -I\langle m_z^2\rangle/4$ (here, $I$ is a Stoner exchange interaction, $\langle m_z^2\rangle$ - square of the local magnetic moment), and therefore favor the corundum structured (space group *R-3c*) phase, i.e., the phase with high local magnetic moments. As a result, the intermediate DPv phase, with site-selective electronic and magnetic properties, is stabilized at intermediate pressures of about 50-60 GPa. This is presumably an electronic phase transition that results in the appearance of (at least) two electronically/magnetically different sublattices of Fe-cations; i.e., it always leads to a structural transformation, due to electron-lattice coupling. We point out that similar behavior only associated with the site-selective HS-LS transition is found to occur in a two-orbital Hubbard model with crystal-field splitting[43]. Meanwhile, during decompression we observe the "pure" electronic transition back to a conventional Mott insulating state within the DPv phase, with decoupling of the electronic and lattice (crystal-structure) degrees of freedom. This behavior is similar to what happens in Fe-bearing bridgmanite ($MgSiO_3$-peroskite) where only half of the $Fe^{3+}$ (those in the *B* site) undergoes a HS to LS transition under pressure, while $Fe^{3+}$ in the *A* site remains in the HS state up to at least 100 GPa[44,45]. We note, however, the important difference that in $Fe_2O_3$ upon decompression the LS-HS transition is accompanied by a transformation to the conventional Mott insulating phase.

Upon compression of $Fe_2O_3$ above 62 GPa, we observe a further increase in the abundance of a non-magnetic component of the $Fe^{3+}$ cations, which is presumably caused by the onset of the Fe *A*-sites of the DPv structure transforming into a metallic non-magnetic state. This results in a structural transition from the DPv to the *Aba2* phase, corresponding to completion of the electronic transition. Thus, starting from a structure with a single crystallographic site for $Fe^{3+}$ cations, $Fe_2O_3$ transforms into an intermediate structure containing multiple Fe-sites, half of which are metallic, before the tendency of $Fe^{3+}$ to metallize results in a 2nd structural transition to the metallic *Aba2* phase, which again has a single octahedral site. This demonstrates the real level of complexity of electronic and structural transformations that can arise in strongly correlated transition-metal compounds undergoing a Mott insulator-to-metal transition.

Our results suggest that the concept of a site-selective Mott transition may be broadly applicable to correlated-electron materials: in particular, in those with a corundum crystal structure, as in the case of $Fe_2O_3$. For example in $Mn_2O_3$, corundum-type ε-$Mn_2O_3$ (and below $T = 1200$ K cubic α- $Mn_2O_3$) transforms upon compression to a distorted perovskite structure[46], and this structural transition concurs with an IM transition[47]. Similar electronic transformations could be expected in other sesquioxides, e.g. $Cr_2O_3$[48], $Ti_2O_3$[49], and in materials with a complex crystal structure (or that acquire a complex structure under pressure) containing TM cations in different coordination polyhedra: for example, in magnetite[50] or Fe-bearing bridgmanite[44,45]. Thus, such effect(s) can occur in crystalline oxides comprising Earth and planetary mantles. Indeed, the major components of Earth's lower mantle – bridgmanite and ferropericlase – contain either or both ferrous and ferric iron, changes of the electronic state in such materials likely affect the properties of the our planet's deep interior[51,52].

**Materials and Methods**

**Samples**

The $Fe_2O_3$ powder (99.5% pure) used in this study is commercially available from Riedel-de Haën. For Mössbauer studies, 30% enriched $^{57}Fe_2O_3$ was used. For single-crystal X-ray diffraction, the same hematite single crystals described elsewhere[32,53] were used.

**Experimental Methods**

Custom 4-pin diamond anvil cells (DACs) made at Tel-Aviv University[54] and Bayerisches Geoinstitut, with anvil culet diameters of 250 or 200 μm and Re gaskets, were used to induce high pressure. Neon was used as a pressure-transmitting medium. Pressure was determined using the ruby R1 fluorescence line as a pressure marker, and the Ne unit-cell volume in the case of XRD studies.

*$^{57}Fe$ Mössbauer* studies were performed up to 80 GPa using a 10 mCi $^{57}Co$ (Rh) point source in a variable temperature (5 – 300 K) cryostat. Spectra were analyzed using a Spin-Hamiltonian fitting

program[55] from which the isomer shift (IS), the quadrupole splitting (QS), the hyperfine field ($H_{hf}$) and the relative abundances of the spectral components were deduced. The reported velocity is with respect to α-Fe. The spectrum at 79 GPa and 4 K was collected using energy-domain synchrotron Mössbauer spectroscopy carried out at the beamline ID18 at ESRF[56]. This spectrum was collected with the source at RT and, therefore, is affected by the 2[nd] order Doppler shift.

*Electrical resistance* measurements were performed up to 90 GPa. The Re gasket was covered with an insulating layer of an $Al_2O_3$-NaCl mixture (3:1 atomic ratio), which also serves as the pressure medium. Platinum foils with a thickness of 5-7 μm were cut in triangular form and used as electrical probes for resistance measurements. The foils were connected to copper leads, at the base of the diamond anvil, using a silver epoxy. Resistance was measured as a function of pressure and temperature using a standard four-probe method in a custom-made cryostat. At each temperature, the voltage was measured as a function of a series of applied currents, for determining the resistance from the obtained slope. Pressure was measured by ruby fluorescence both before and after each measurement.

*Powder X-ray diffraction* experiments were performed at the Extreme Conditions Beamline (ECB) P02.2 at PETRA III, Hamburg, Germany (λ = 0.28953 Å) in angle-dispersive mode with patterns collected using a Perkin Elmer (PE) flat panel detector and integrated using the FIT2D program[57]. The results were analyzed by Rietveld refinement using the GSAS package[58] and EXPGUI[59].

Details of the single-crystal diffraction experiments are given in references 32, 33 and 37.

The reported uncertainties are given according to the standard errors obtained from the respective software used for fitting the data.

## Theoretical Methods

We calculate the electronic structure and spin state of paramagnetic $Fe_2O_3$ using the DFT+DMFT computational approach (DMFT: dynamical mean-field theory). The DFT+DMFT method[36] describes the quantum dynamics of the many-electron problem exactly (neglecting non-local effects) and allows one to include the effect of electronic correlations on the electronic properties and lattice structure of correlated materials. For the partially filled Fe *3d* and O *2p* orbitals we construct a basis set of atomic-centered symmetry-constrained Wannier functions[60]. In order to enhance localization of the Fe *3d* Wannier orbitals, the O *2p* orbitals were constructed using Wannier functions defined over the full energy range spanned by the *p-d* band complex; the localized Fe *3d* orbitals are constructed using the Fe *3d* band set. To solve the realistic many-body problem, we employ the continuous-time hybridization-expansion quantum Monte-Carlo algorithm[61]. In the DPv phase, correlations in the Fe *3d* bands of the structurally distinct Fe *A* and Fe *B* sites are treated using a cluster expansion of the DFT+DMFT approach. The calculations are performed in the paramagnetic state at temperature $T = 390$ K. We use the average Coulomb interaction $U = 6$ eV and Hund's exchange $J = 0.86$ eV for the Fe *3d* shell as was estimated previously[35]. The Coulomb interaction is treated in the density-density approximation. The spin-orbit coupling is neglected in these calculations. We employ the fully localized double-counting correction, evaluated from the self-consistently determined local occupancies, to account for the electronic interactions already described by DFT. The spectral functions were computed using the maximum entropy method. Further technical details about the method used can be found in Leonov et al.[62].

**Acknowledgments:**

We would like to thank Prof. D. Vollhardt and Dr. W. Xu for valuable discussions, Dr. R. Rüffer for experimental assistance with the facilities of the ID18 beam line at ESRF, Dr. A. Kurnosov for loading of DACs with Ne gas. This research was supported in part by Israeli Science Foundation Grant #1189/14. I.L. acknowledges support by the Deutsche Forschungsgemeinschaft through Transregio TRR 80 and the Ministry of Education and Science of the Russian Federation in the framework of Increase Competitiveness Program of NUST 'MISIS' (K3-2016-027), implemented by a governmental decree dated 16th of March 2013, N 211. I.A.A. is grateful for the support from the Swedish Research Council

(VR grant 2015-04391), the Swedish Government Strategic Research Areas in Materials Science on

Functional Materials at Linköping University (Faculty Grant SFO-Mat-LiU No 2009 00971) and the

Swedish e-science Research Centre (SeRC), as well as from the Ministry of Education and Science of

the Russian Federation in the framework of grant No. 14.Y26.31.0005 and Increase Competitiveness

Program of NUST "MISIS" (No. K2-2016-013).

**Author contributions:** E.G., S.L., Z.K., and G.K.R. conducted the experiments; I.L. and I.A.A.

performed the theoretical analysis; G.K.R., R.J., E.G., and I.L. wrote the manuscript and all the authors contributed to the interpretation of the data and to the writing of the final manuscript; G.K.R., I.A.A.

L.D. and R.J. conceived and supervised the project.

**Competing interests:** The authors declare that they have no competing interests.

**Data and materials availability:** Data is available upon request to the corresponding authors.

## References

1. Mott NF (1990) *Metal-Insulator Transitions*. (Taylor & Francis, London).

2. Imada M, Fujimori A, Tokura Y (1998) Metal-insulator transitions. *Rev Mod Phys* 70:1039–1263.

3. Si Q, Abrahams E (2008) Strong Correlations and Magnetic Frustration in the High Tc Iron Pnictides. *Phys Rev Lett* 101:076401–076404.

4. de'Medici L, Gianluca G, Capone M (2014) Selective Mott Physics as a Key to Iron Superconductors. *Phys Rev Lett* 112:177001–177005.

5. Yi M, et al. (2015) Observation of universal strong orbital-dependent correlation effects in iron chalcogenides. *Nat Commun* 6:7777–7783.

6. Vojta M (2010) Orbital-selective Mott transitions: Heavy fermions and beyond. *J Low Temp Phys* 161:203–232.

7. Arita R, Held K, Lukoyanov AV, Anisimov VI (2007) Doped Mott Insulator as the Origin of Heavy-Fermion Behavior in $LiV_2O_4$. *Phys Rev Lett* 98:166402–166406.

8. Rozenberg GK, Xu W, Pasternak MP (2014) The Mott insulators at extreme conditions; structural consequences of pressure-induced electronic transitions. *Zeitschrift für Krist – Cryst Mater* 229:210–222.

9. Leonov I, Anisimov VI, Vollhardt D (2015) Metal-insulator transition and lattice instability of paramagnetic $V_2O_3$. *Phys Rev B* 91:195115–195119.

10. Kuneš J, Lukoyanov AV, Anisimov VI, Scalettar RT, Pickett WE (2008) Collapse of magnetic moment drives the Mott transition in MnO. *Nature Mater* **7**:198-202.

11. Leonov I (2015) Metal-insulator transition and local-moment collapse in FeO under pressure. *Phys Rev B* 92:085142–085147.

12. Lyubutin IS, Ovchinnikov SG, Gavriliuk AG, Struzhkin VV (2009) Spin-crossover-induced Mott transition and the other scenarios of metallization in $3d^n$ metal compounds. *Phys Rev B* 79:085125–085130.

13. Greenberg E, et al. (2013) Mott transition in $CaFe_2O_4$ at around 50 GPa. *Phys Rev B* 88:214109-214113.

14. A. G. Gavriliuk, V. V. Struzhkin, I. S. Lyubutin, S. G. Ovchinnikov, M. Y. Hu, and P. Chow (2008), Phys. Rev. B **77**, 155112.

15. Anisimov VI, Nekrasov IA, Kondakov DE, Rice TM, Sigrist M (2002) Orbital-selective Mott-insulator transition in $Ca_{2-x}Sr_xRuO_4$. *Eur Phys J B* 25:191–201.

16. Liebsch A (2003) Mott Transitions in Multiorbital Systems. *Phys Rev Lett* 91:226401–226404.

17. Koga A, Kawakami N, Rice TM, Sigrist M (2004) Orbital-selective Mott transitions in the degenerate Hubbard model. *Phys Rev Lett* 92:216402–216405.

18. de'Medici L, Georges A, Biermann S (2005) Orbital-selective Mott transition in multiband systems: Slave-spin representation and dynamical mean-field theory. *Phys Rev B* 72:205124–205138.

19. F. van der Woude, Phys. Status Solidi B **17**, 417 (1966).

20. A. Fujimori, M. Saeki, N. Kimizuka, M. Taniguchi and S. Suga, (1986) Phys. Rev. B **34**, 7318.

21. R. G. McQueen and S. P. Marsh, in *Handbook in Physical Constants*, edited by S. P. Clark, revised edition (Geological Society of America, Inc., 1966), p. 153.

22. Olsen JS, Cousins CSG, Gerward L, Jhans H, Sheldon BJ (1991) A study of the crystal structure of $Fe_2O_3$ in the pressure range up to 65 GPa using synchrotron radiation. *Phys Scr* 43:327–330.

23. Rozenberg GK, et al. (2002) High-pressure structural studies of hematite $Fe_2O_3$. *Phys Rev B* 65: 064112–064119.

24. Suzuki T, et al. (1985) *Solid State Physics under Pressure*, eds. Minomura S (Terra Scientific Publishing Company), pp. 149-154.

25. Pasternak M, et al. (1999) Breakdown of the Mott-Hubbard State in $Fe_2O_3$: A First-Order Insulator-Metal Transition with Collapse of Magnetism at 50 GPa. *Phys Rev Lett* 82:4663–4666.

26. T. Yagi and S. Akimoto, in High Pressure Research in Geophysics, edited by S. Akimoto and M. H. Manghnani (Center Academic Publishers, Tokyo, 1982), p. 81.

27. Syono Y, et al. (1984) Mössbauer study on the high pressure phase of $Fe_2O_3$. *Solid State Commun* 50: 97–100.

28. Nasu S, Kurimoto K, Nagatomo S, Endo S, Fujita FE (1986) $^{57}Fe$ Mössbauer study under high pressure; ε-Fe and $Fe_2O_3$. *Hyperfine Interact* 29:1583–1586.

29. Badro J, et al. (1999) Magnetism in FeO at Megabar Pressures from X-Ray Emission Spectroscopy. *Phys Rev Lett* 83:4101–4104.

30. Wang S, et al. (2010) High pressure evolution of $Fe_2O_3$ electronic structure revealed by X-ray absorption. *Phys Rev B* 82:144428–144432.

31. Reid AF, Ringwood AE (1969) High-pressure scandium oxide and its place in the molar volume relationships of dense structures of $M_2X_3$ and $ABX_3$ type. *J Geophys Res* 74:3238–3252.

32. Bykova E, et al. (2013) Novel high pressure monoclinic $Fe_2O_3$ polymorph revealed by single-crystal synchrotron X-ray diffraction studies. *High Press Res* 33:534–545.

33. Bykova E, et al. (2016) Structural complexity of simple $Fe_2O_3$ at high pressures and temperatures. *Nat Commun* 7:10661–10666.

34. Badro J, et al. (2002) Nature of the High-Pressure Transition in $Fe_2O_3$ Hematite. *Phys Rev Lett* 89: 205504–205507.

35. Kuneš J, Korotin DM, Korotin MA, Anisimov VI, Werner P (2009) Pressure-Driven Metal-Insulator Transition in Hematite from Dynamical Mean-Field Theory. *Phys Rev Lett* 102:146402–146405.

36. Georges A, Kotliar G, Krauth W, Rozenberg MJ (1996) Dynamical mean-field theory of strongly correlated fermion systems and the limit of infinite dimensions. *Rev Mod Phys* 68:13–125.

37. Bykova E (2015) Single-crystal X-ray diffraction at extreme conditions in mineral physics and material sciences. (Universitaet Bayreuth, Bayreuth). https://epub.uni-bayreuth.de/2124/1/Thesis_Bykova_final_version.pdf.

38. Machavariani GYu, Pasternak MP, Rozenberg GKh (2000) High pressure Metallization of Hematite, in *Science and Technology of High Pressure V2, Proceedings of the International Conference on High Pressure Science and Technology (AIRAPT-17)*, AIRAPT, Honolulu, Hawii, 25-30 July 1999 (Universities Press, Hyderabad, India), pp. 562.

39. Kondo J, (1964) Resistance Minimum in Dilute Magnetic Alloys. *Prog Theor Phys* 32:37–49.

40. Park H, Millis AJ, Marianetti CA (2012) Site-Selective Mott Transition in Rare-Earth-Element Nickelates. *Phys Rev Lett* 109:156402–156406.

41. Subedi A, Peil OE, Georges A (2015) Low-energy description of the metal-insulator transition in the rare-earth nickelates. *Phys Rev B* 91:075128–075143.

42. Merlini M, et al. (2010) $Fe^{3+}$ spin transition in $CaFe_2O_4$ at high pressure. *Am Mineral* 95:200–203.

43. J. Kuneš and V. Křápek (2013). Disproportionation and Metallization at Low-Spin to High-Spin Transition in Multiorbital Mott Systems. Phys. Rev.Lett. **106**, 256401.

44. K. Catalli, S.-H. Shim, V. B. Prakapenka, J. Zhao, W. Sturhahn, P. Chow, Y. Xiao, H. Liu, H. Cynn, W. J. Evans (2010). Spin state of ferric iron in $MgSiO_3$ perovskite and its effect on elastic properties. Earth Planet. Sci. Lett. **289** 68-75.

45. H. Hsu, P. Blaha, M. Cococcioni, and R. M. Wentzcovitch (2011). Spin-State Crossover and Hyperfine Interactions of Ferric Iron in $MgSiO_3$ Perovskite. Phys. Rev. Lett. **106**, 118501.

46. Ovsyannikov SV, et al. (2013) Perovskite-like $Mn_2O_3$: A Path to New Manganites, Angew. Chem. Int. Ed. **52**, 1494

47. Hong F, Yue B, Hirao N, Liu Z., Chen B (2017) Significant improvement in $Mn_2O_3$ transition metal oxide electrical conductivity via high pressure, Scientific Rep. 7:44078

48. Shim S-H, Duffy TS, Jeanloz R, Yoo C-S, Iota V (2004) Raman spectroscopy and x-ray diffraction of phase transitions in $Cr_2O_3$ to 61 GPa. *Phys Rev B* 69:144107–144118.

49. Ovsyannikov SV, et al. (2013) High-pressure behavior of structural, optical, and electronic transport properties of the golden $Th_2S_3$-type $Ti_2O_3$. *Phys Rev B* 88:184106–184120.

50. E. Greenberg, W. M. Xu, M. Nikolaevsky, E. Bykova, G. Garbarino, K. GLazyrin, D. G. Merkel, L. Dubrovinsky, Mp. P. Pasternak and G. Kh. Rozenberg (2017). High-pressure magnetic, electronic and structural properties of $M$$Fe_2O_4$ ($M$= Mg, Zn, Fe) ferric spinels. Phys. Rev. B **95**, 195150-1-13.

51. Potapkin V, et al. (2013) Effect of iron oxidation state on the electrical conductivity of the Earth's lower mantle. *Nat Commun* 4:1427–1433.

52. Mao, Z., Lin, J., Yang, J., Inoue, T. & Prakapenka, V. B. Effects of the $Fe^{3+}$ spin transition on the equation of state of bridgmanite. *Geophys. Res. Lett.* **42,** 4335–4342 (2015).

53. Schouwink P, et al. (2011) High-pressure structural behavior of α-$Fe_2O_3$ studied by single-crystal X-ray diffraction and synchrotron radiation up to 25 GPa. *Am Mineral* 96:1781–1786.

54. Machavariani GYu, Pasternak MP, Hearne GR, Rozenberg GK (1998) A multipurpose miniature piston-cylinder diamond-anvil cell for pressures beyond 100 GPa. *Rev Sci Instrum* 69:1423–1425.

55. Prescher C, McCammon C, Dubrovinsky L, (2012) Moss A: a program for analyzing energy-domain Mössbauer spectra from conventional and synchrotron sources. *J Appl Crystallogr* 45:329–331.

56. Potapkin V, et al. (2012) The $^{57}Fe$ Synchrotron Mössbauer Source at the ESRF. *J Synchrotron Radiat* 19:559–569.

57. Hammersley AP (1998) *ESRF Internal Report*, ESRF98HA01T, FIT2D V9.129 Reference Manual V3.1 (ESRF, Grenoble, France).

58. Larson AC, Von Dreele RB (1994) General Structure Analysis System (GSAS), Los Alamos National Laboratory, LAUR 86–748.

59. Brian HT (2001) EXPGUI, a graphical user interface for GSAS. *J Appl Cryst* 34:210–213.

60. Anisimov VI, et al. (2005) Full orbital calculation scheme for materials with strongly correlated electrons. Phys Rev B 71:125119–125134.

61. Gull E, Millis AJ, Lichtenstein AI, Rubtsov AN, Troyer M, Werner P (2011) Continuous-time Monte Carlo methods for quantum impurity models. *Rev Mod Phys* 83:349–404.

62. Leonov I, Korotin Dm, Binggeli N, Anisimov VI, Vollhardt D (2010) Computation of correlation-induced atomic displacements and structural transformations in paramagnetic $KCuF_3$ and $LaMnO_3$. *Phys Rev B* 81:075109–075119.




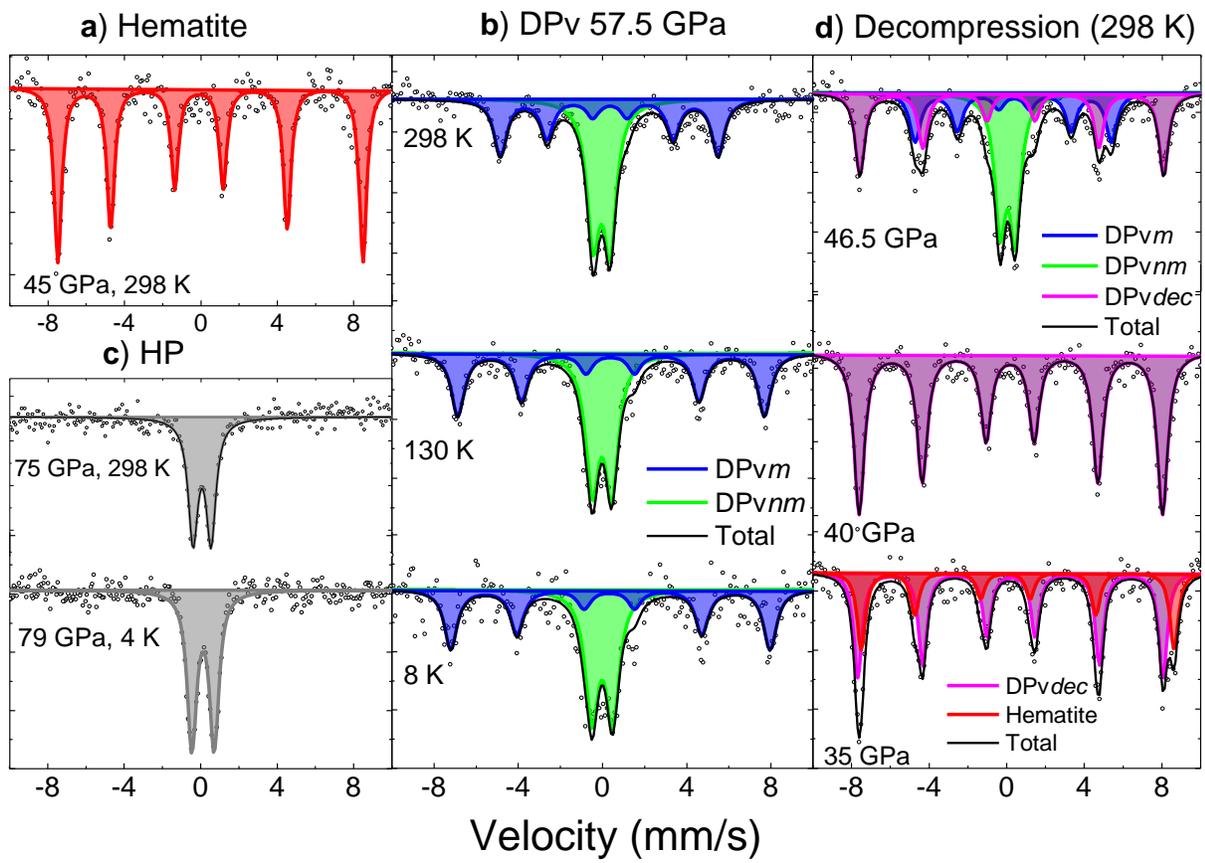

a) Hematite

45 GPa, 298 K

b) DPv 57.5 GPa

298 K

DPv*m*
DPv*nm*
Total

130 K

DPv*m*
DPv*nm*
Total

8 K

c) HP

75 GPa, 298 K

79 GPa, 4 K

d) Decompression (298 K)

46.5 GPa

DPv*m*
DPv*nm*
DPv*dec*
Total

40 GPa

35 GPa

DPv*dec*
Hematite
Total

Velocity (mm/s)



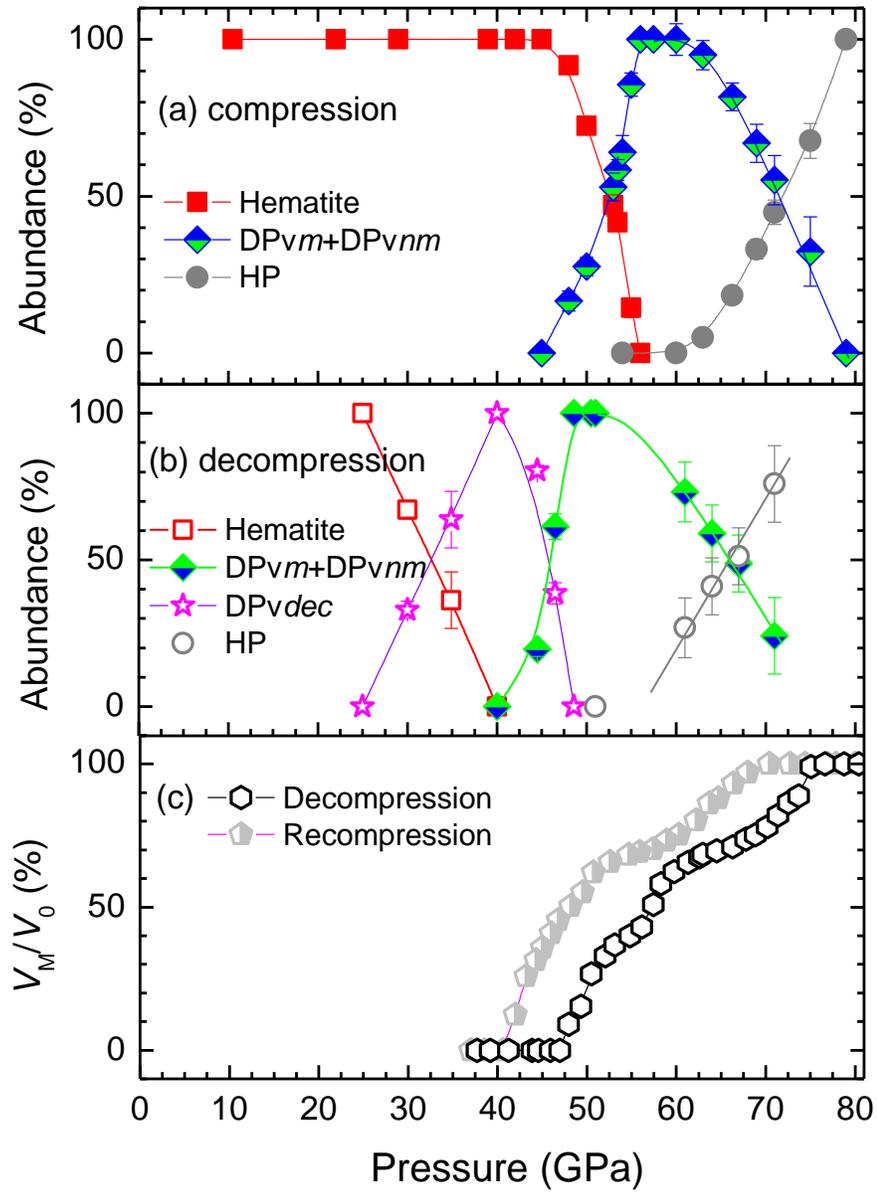



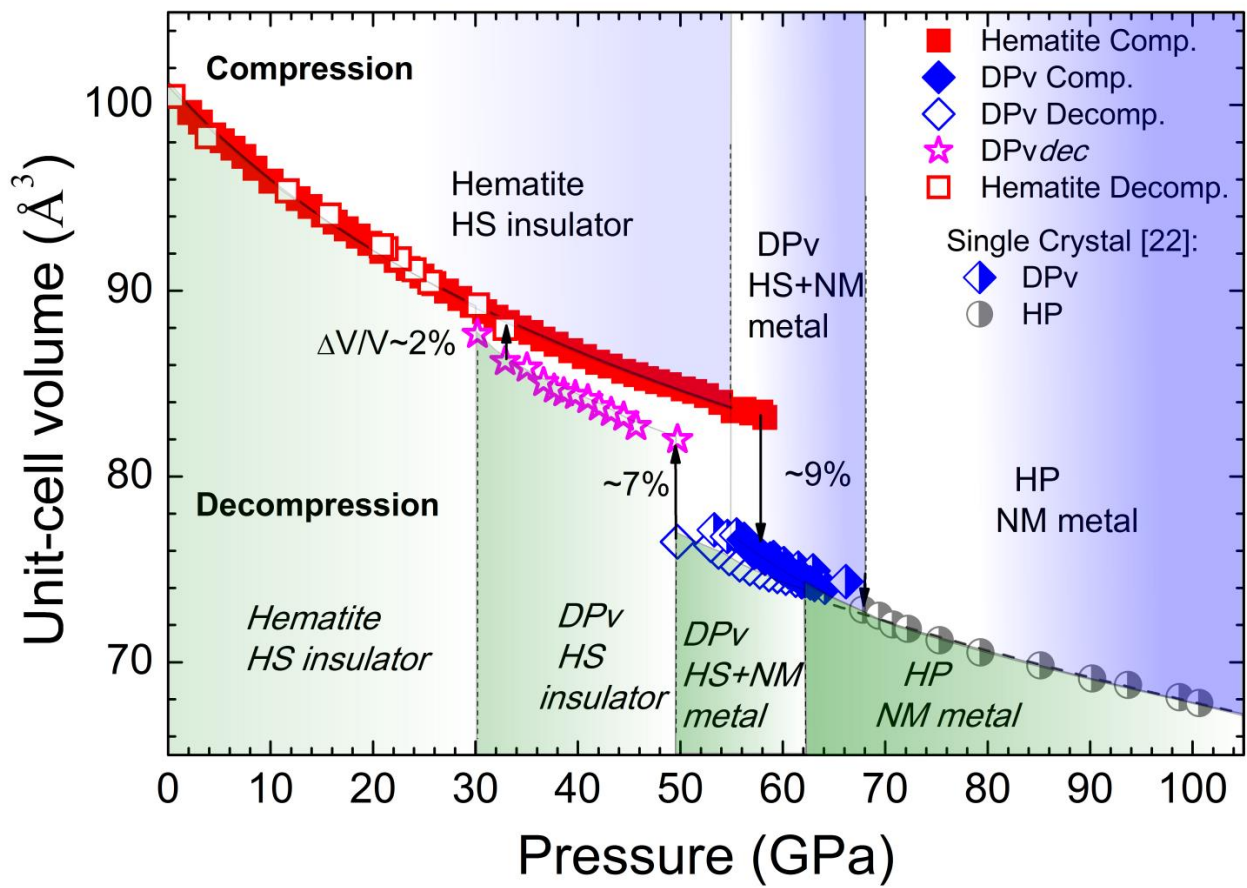



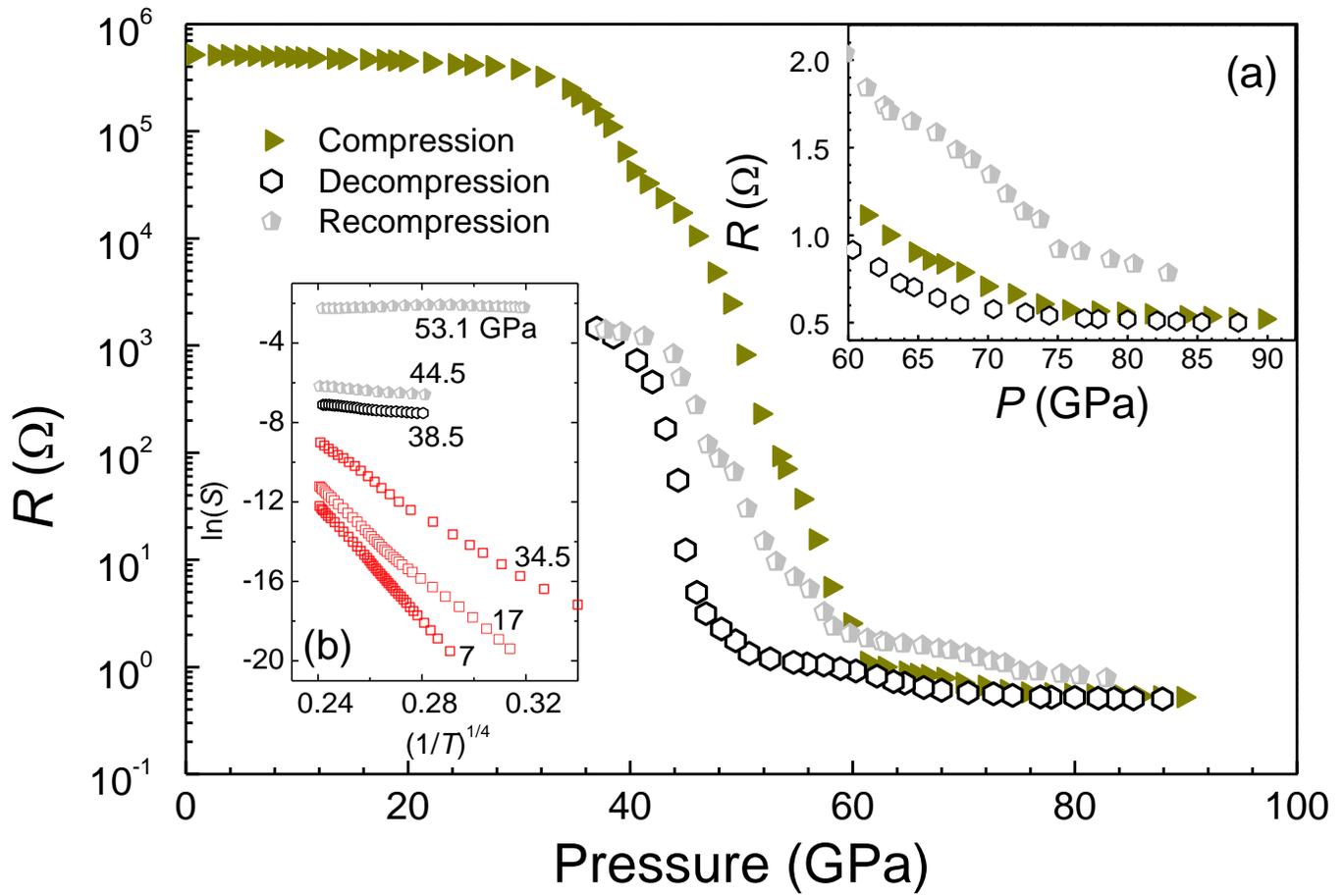



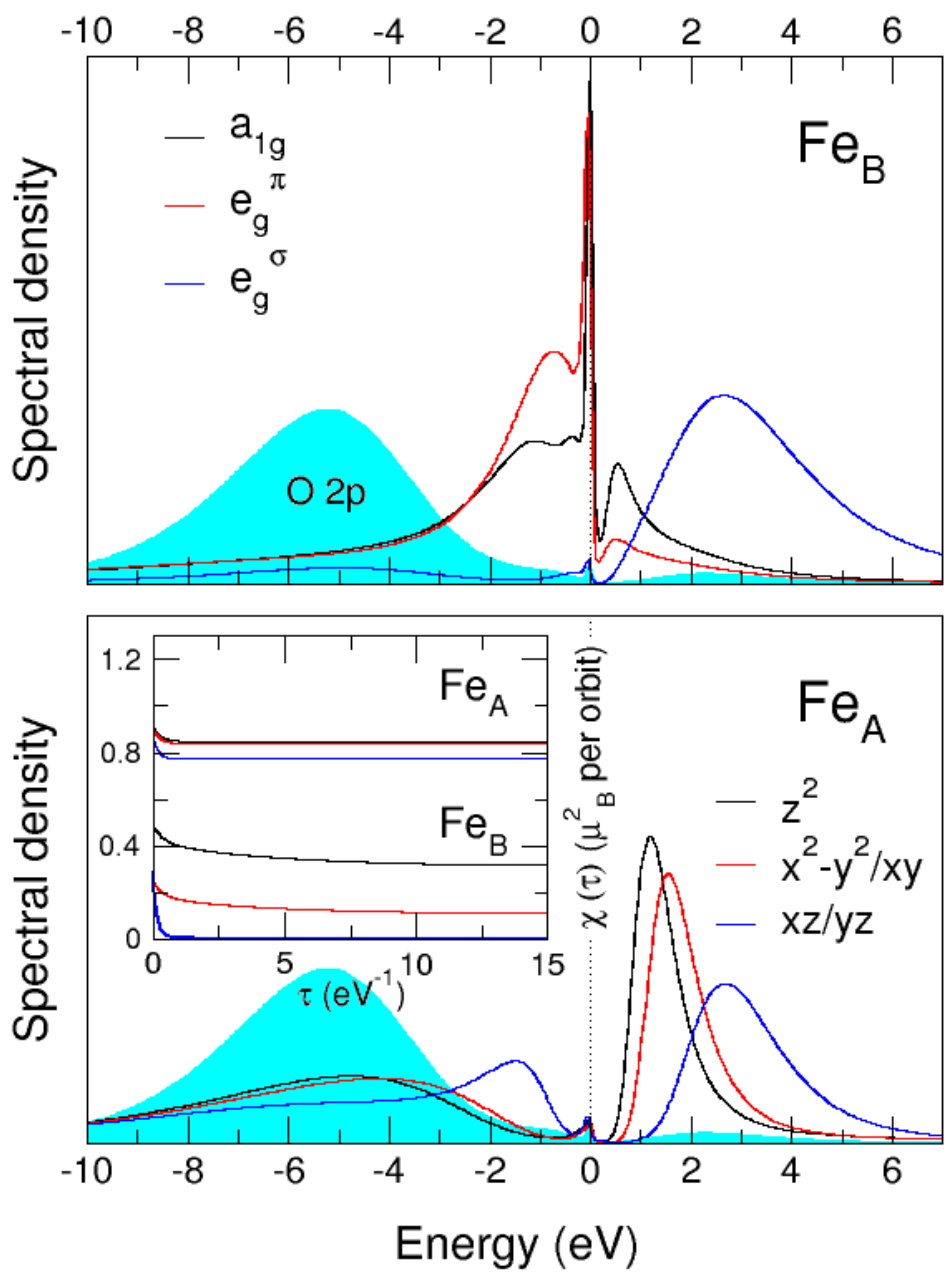





**Figure 1.** $^{57}$Fe Mössbauer spectra of $Fe_2O_3$ at various pressures and temperatures: at room temperature (**a, d**) and reduced temperatures (**b, c**) upon compression (**a, b, c**) and decompression (**d**). Red, blue, green and magenta lines and shaded areas represent fits to the hematite, magnetic DPv (DPv$_m$), non-magnetic DPv (DPv$_{nm}$) and DPv$_{dec}$ components; the black solid line is the sum of all components. The high-pressure phase (HP) component is shown in solid grey. Upon compression to ~48 GPa (**a**) the only observed spectral component is that of the $Fe^{3+}$ high-spin state. Above 48 GPa the intermediate pressure DPv phase emerges, characterized by two equally abundant components: high-spin DPv$_m$ and non-magnetic DPv$_{nm}$. At $P \geq 75$ GPa the spectra show a single, quadrupole-split component, with no sign of magnetic correlations down to 4 K (**c**, 79 GPa). Upon decompression one can see the onset of the new DPv$_{dec}$ component at 46.5 GPa. At 35 GPa the original hematite component appears, replacing with further decompression the DPv$_{dec}$ component.

**Figure 2.** Pressure evolution of the relative abundances of the Mössbauer spectral components, and the abundance of the metallic phase. The abundances of the Mössbauer spectral components are shown for compression (**a**) and decompression (**b**) as computed from the relative areas of the absorption bands. All lines are to guide the eye. Symbols "■" represent the hematite, "◆" - combined DPv$_m$ (magnetic) and DPv$_{nm}$ (non-magnetic) components, "●" - HP component and "☆" - DPv$_{dec}$ component. Filled and empty shapes mark compression and decompression cycles, respectively. The pressure uncertainties are 1-2 GPa. Note the onset of the new DPv$_{dec}$ component upon decompression below 47 GPa, with significant hysteresis in the reappearance of the DPv and hematite components. Upon recompression from 25 GPa (not shown), at 50 GPa, the DPv$_{nm}$ and DPv$_m$ components appear again with a relative



abundance in agreement with the compression trend. The relative abundance of the metallic phase is shown for decompression and recompression (from 37 GPa) (**c**) as derived from the room-temperature measurements of electrical resistance (see Supporting Info). Open hexagons and half-filled pentagons denote successive decompression and recompression, respectively: note the two distinct steps in the $V_M/V_0(P)$ dependence. The solid curves are only guides for the eye.

**Figure 3.** Pressure dependence of the unit-cell volume of $Fe_2O_3$ at room temperature, with volume shown normalized to 2 formula units of $Fe_2O_3$. Solid and open symbols indicate compression and decompression (powder), respectively, and half-filled symbols correspond to the single-crystal compression experiments. Solid and dashed lines are fits for the hematite and HP (space group *Aba*2) phases using the second-order Birch-Murnaghan equation of state (EOS). The zero-pressure bulk modulus, unit-cell volume and bulk modulus first derivative obtained for the $R\bar{3}c$ structure are: $K_0 = 197.6(6)$ GPa, $V_0 = 100.58(3)$ Å$^3$, $K' = 4$ (fixed); when fitting data to the third-order EOS: $K_0 = 187(3)$ GPa, $V_0 = 100.75(5)$ Å$^3$, $K' = 4.5(2)$. Note the two structural transitions upon compression ($R\bar{3}c \rightarrow$ DPv $\rightarrow$ *Aba*2), while upon decompression an additional isostructural transition is observed within the DPv phase around 50 GPa. The structural transition back to the hematite phase occurs only below 35 GPa and is accompanied by ~ 2% volume increase. The volume error bars are within the symbol sizes. The pressure uncertainties are ~0.1 GPa.

**Figure 4.** Pressure dependence of electrical resistance at 298 K, the solid triangles, open hexagons and half-filled pentagons showing data recorded upon compression, decompression and successive recompression, respectively. Our results for 60 – 90 GPa are shown magnified in the inset (a) in order to emphasize the changes in the pressure dependence at 70 – 75 GPa. In the inset (b) we show the temperature dependence of electrical conductance $S$ of $Fe_2O_3$ at various pressures. Measurements on the



DPv phase were performed during decompression to 37 GPa, and following recompression (empty hexagons and half-filled pentagons, respectively). Measurements of the hematite phase were collected during a separate decompression cycle to ambient pressure (symbols "□").[28] Note the linear relationship of $\ln(S)$ versus $T^{-1/4}$ for both insulating phases, typical for many transition-metal compounds near the insulator-metal transition [Refs. 1, 2 and references therein]. The change in sign of the slope $d\ln(S)/dT$ documented at ~53 GPa is the signature of metallic conductivity.

**Figure 5.** Spectral function of paramagnetic DPv $Fe_2O_3$ calculated by DFT+DMFT at temperature $T = 390$ K, based on crystal structure parameters taken from the X-ray diffraction results at ~54 GPa. The contributions from Fe *3d* (black, red and blue curves) and O *2p* (blue-shaded area) orbitals are shown as a function of energy normalized to the Fermi energy (i.e., $E_F = 0$). The calculations show the existence of a site-selective Mott phase, in which the *3d* electrons of only half of the Fe sites (octahedral *B* sites: *top*) are metallic with $a_{1g}$ and $e_g^\pi$ curves showing significant amplitude below the Fermi energy, while the other (*A* sites: *bottom*) remain insulating as indicated by the amplitude for the Fe *3d* curves being almost entirely above the Fermi energy. Our result for the local magnetic moment of the Fe *A* sites is ~4.63 $\mu_B$, and for the *B* sites is ~0.89 $\mu_B$. Inset: Local spin-spin correlation function $\chi(\tau) = \langle m_z(\tau)m_z(0)\rangle$ calculated by DFT+DMFT for paramagnetic DPv $Fe_2O_3$ at $T = 390$ K gives the intra-orbital Fe *3d* contributions. While the Fe *A* electrons are localized to form fluctuating moments ($\chi \approx 1$), the Fe *B* electrons show itinerant magnetic behavior ($\chi << 1$), thereby revealing a transition from localized to site-selective itinerant moments for $Fe_2O_3$ under pressure.





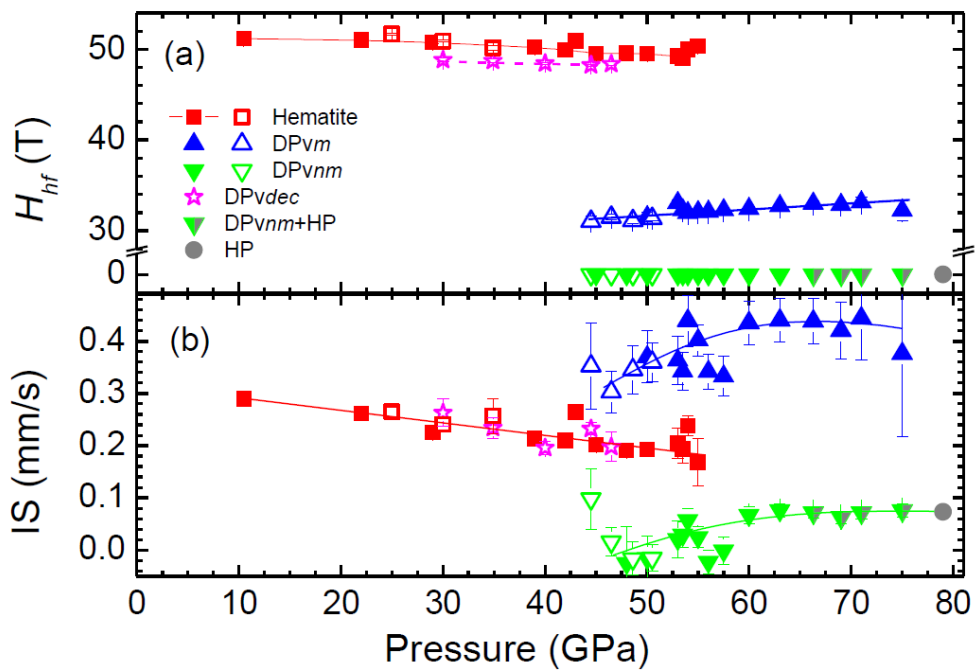

**SI Figure 1.** Hyperfine parameters of $Fe_2O_3$ obtained from $^{57}Fe$ Mössbauer spectroscopy. The pressure dependence of the hyperfine field (**a**) and of the isomer shift (**b**) at 298 K. All lines are to guide the eye. Symbols "■" represent the hematite, "▲" and "▼"- the DPv$_m$ and DPv$_{nm}$ components, respectively, "●" - HP component and "★" - DPv$_{dec}$ component (double-shaded triangles represent the combined indistinguishable DPv$_{nm}$ and HP components). Filled and empty shapes mark compression and decompression cycles, respectively. The pressure uncertainties are 1-2 GPa.

Note the abrupt change in the hyperfine field and of the isomer shift values at the onset of the intermediate pressure DPv phase at ~50 GPa. The non-magnetic DPv$_{nm}$ component has an isomer shift value significantly lower than that of hematite. The decrease in isomer shift means an increase of $s$-electron density $\rho_s(0)$ at the Fe nucleus[62], which is typical for shorter Fe-O distances. The magnetically ordered component DPv$_m$ is characterized by an increased isomer shift value. The corresponding $H_{hf}$ =32.3(3) and 47.3(3) T at room temperature and 8 K (Figure 1(b)), respectively, suggest for DPv$_m$ component a HS $Fe^{3+}$ state with a significantly reduced Néel temperature compared to that of hematite.



Quadrupole splitting increases from 0.4 to 0.6 mm/s upon pressure increase to 50 GPa and then changes to ~0.8 and $\simeq 0$ mm/s for the $DPv_{nm}$ and $DPv_m$ components, respectively.

It is noteworthy, that upon decompression at 35 GPa the hematite and $DPv_{dec}$ phases have magnetic hyperfine field values of $H_{hf}$=50.2(3) and 48.7(3) T, respectively, creating a clear distinction between them.

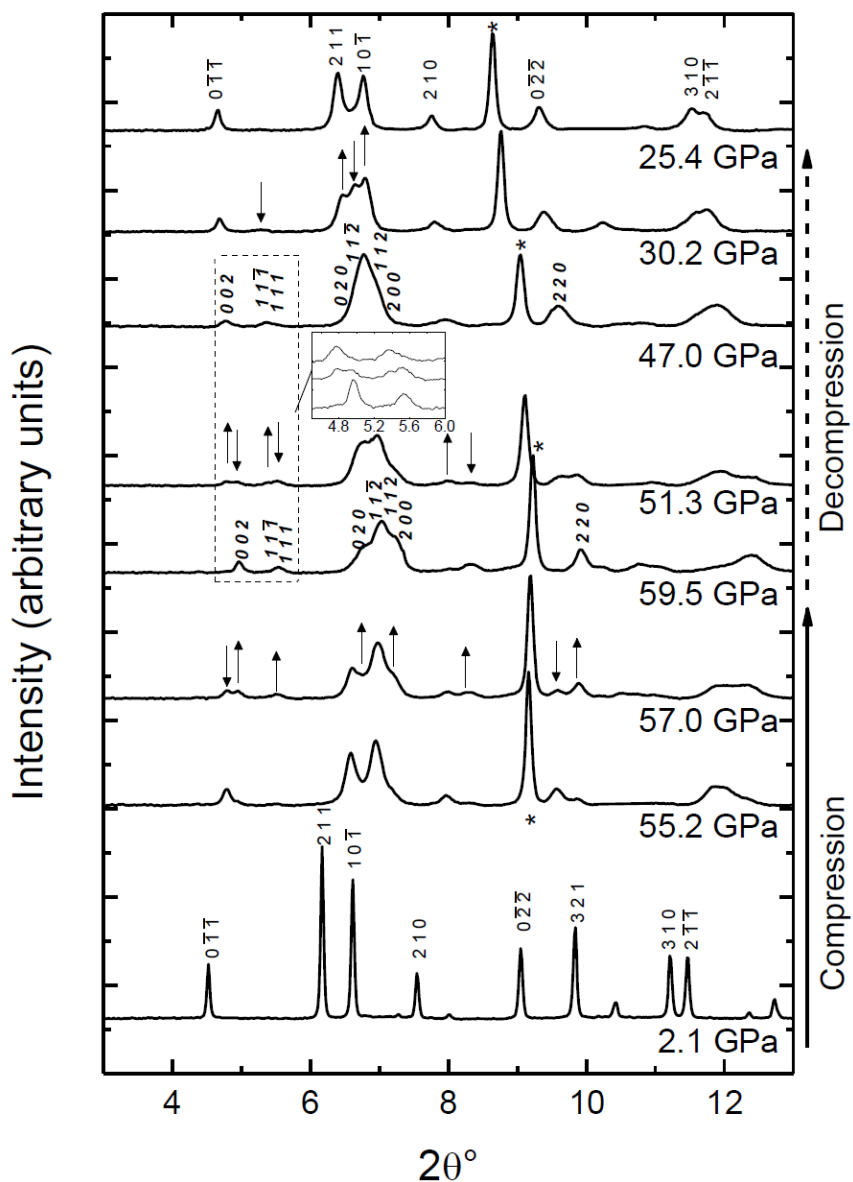

**SI Figure 2.** X-ray powder diffraction patterns of $Fe_2O_3$ at $T$=298 K at various pressures. The DPv intermediate-pressure phase, which first appears at ~53 GPa, is clearly seen at 55.2 GPa.. Upwards



"↑"and downwards "↓" facing arrows represent the increase and decrease of the DPv and hematite phases, respectively. Upon decompression the distinctive doubling of the diffraction peaks of the DPv phase is observed at ~51 GPa, especially (002), (111) and (11-1). A part of the spectrum in the 2θ range of 4.5° – 6.0° is shown in magnification to emphasize the splitting. With further decompression below 35 GPa the peaks of the hematite $R\bar{3}c$ structure appear and at 25 GPa the transition to the original hematite phase is complete. The main diffraction peak of the Ne pressure medium is marked with an asterisk. Italics correspond to the diffraction peaks of the DPv phase.

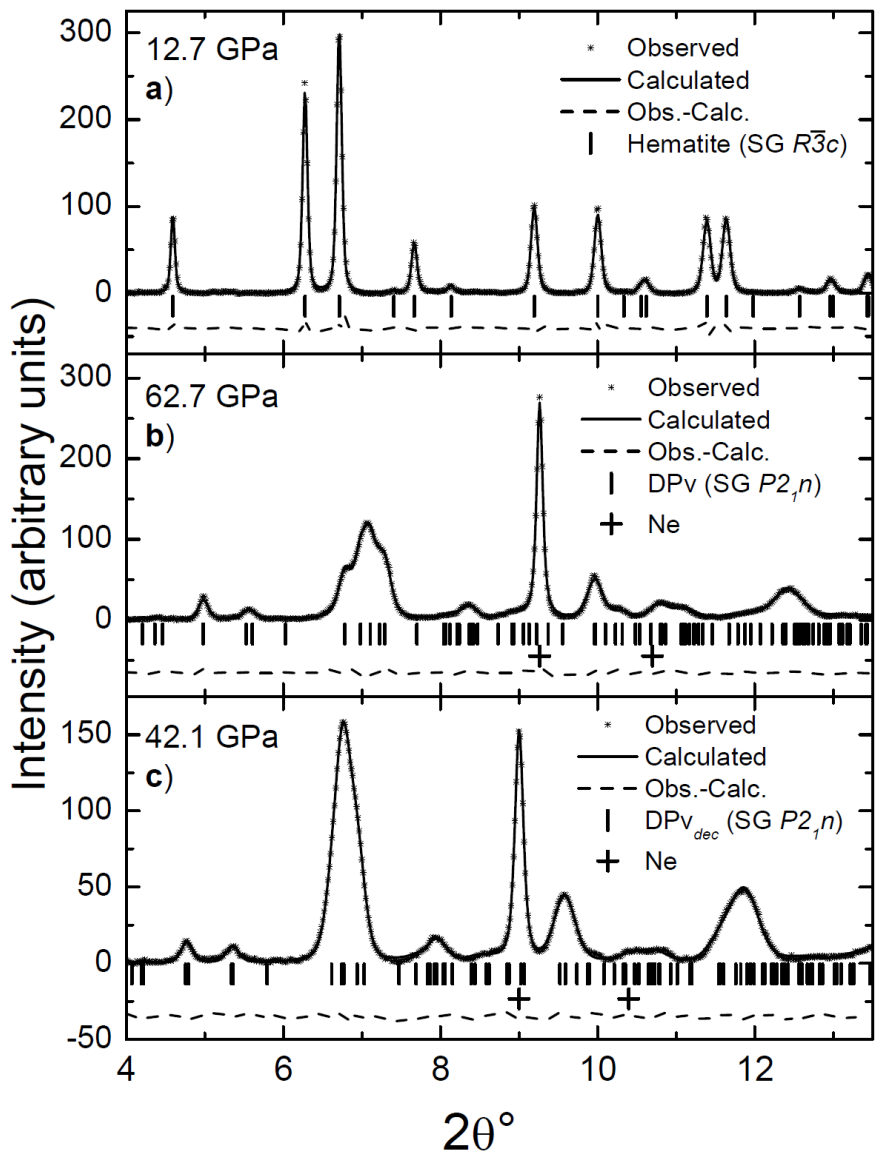



**SI Figure 3.** X-ray diffaction spectra of $Fe_2O_3$ at various pressures. Typical examples of analyzed integrated patterns of XRD spectra collected for hematite and the intermediate-pressure DPv (SG $P2_1/n$) structures upon compression to **a)** 12.7 and **b)** 62.7 GPa, and upon decompression to **c)** 42.1 GPa at RT and the differences between the observed and calculated profiles. Marks show the calculated peak positions. The | and + symbols correspond to $Fe_2O_3$ and Ne pressure medium, respectively. Note that the spectrum at 42.1 GPa, collected upon decompression, is fit well with the $P2_1n$ structure with the reduced monoclinic distortion (marked DPv$_{dec}$).

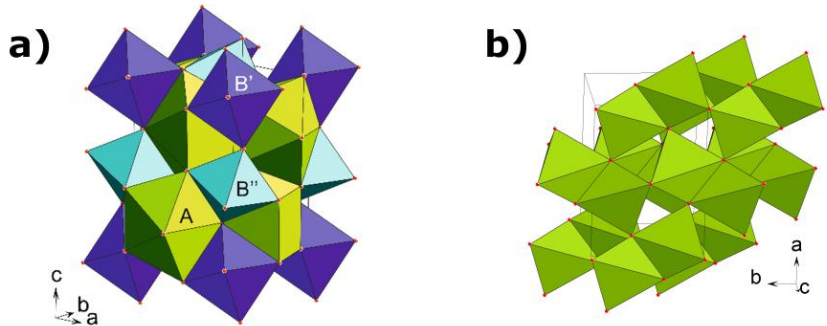

**SI Figure 4.** The high-pressure structures of $Fe_2O_3$. a) Structure of the intermediate-pressure DPv $Fe_2O_3$ phase at 54.6 GPa (triclinic with lattice parameters $a$ = 4.576(6) Å, $b$ = 4.944(2) Å, $c$ = 6.79(2) Å, $\alpha$ = 90.33(7)º, $\beta$ = 89.7(2)º, $\gamma$ = 90.08(5) º). The $A$-positions occupied by HS $Fe^{3+}$, while the octahedral $B$-positions contain nonmagnetic $Fe^{3+}$. b) Structure of the high-pressure phase of $Fe_2O_3$ at 73.8 GPa (space group $Aba$2, $a$=6.524(9) Å, $b$=4.702(3) Å, $c$=4.603(7) Å). In $Aba$2 phase only a singular cation position exists containing nonmagnetic Fe.



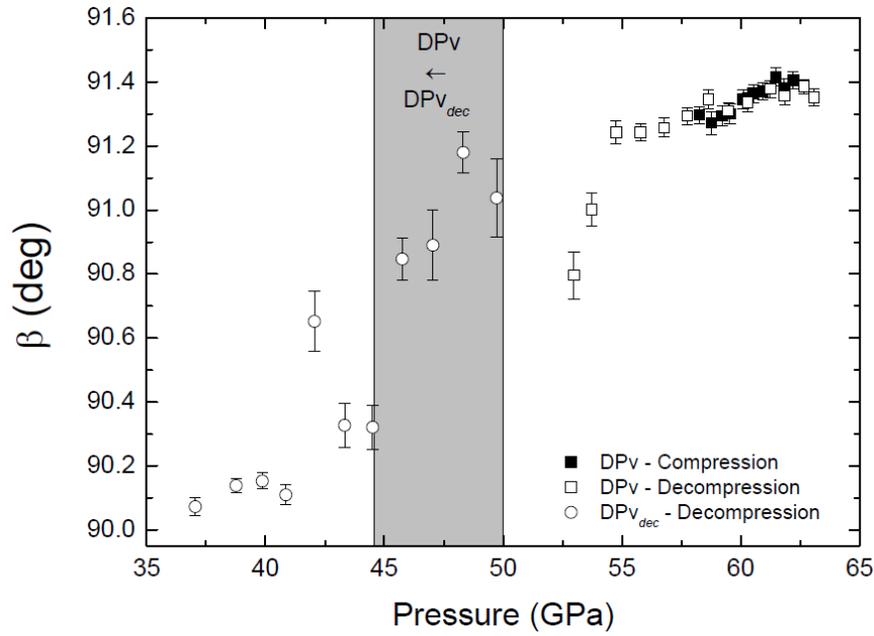

**SI Figure 5.** The monoclinic distortion of the DPv phase of $Fe_2O_3$. Pressure evolution of the β-angle, characteristic of the monoclinic distortion of the unit cell with *P2₁/n* symmetry (DPv structure), upon compression and decompression (solid and open symbols, respectively). Note the significant decrease of the monoclinic distortion coinciding with the reverse electronic transition to the strongly correlated state upon decompression; the *β*-angle changes from 91.4 to 90.1°.

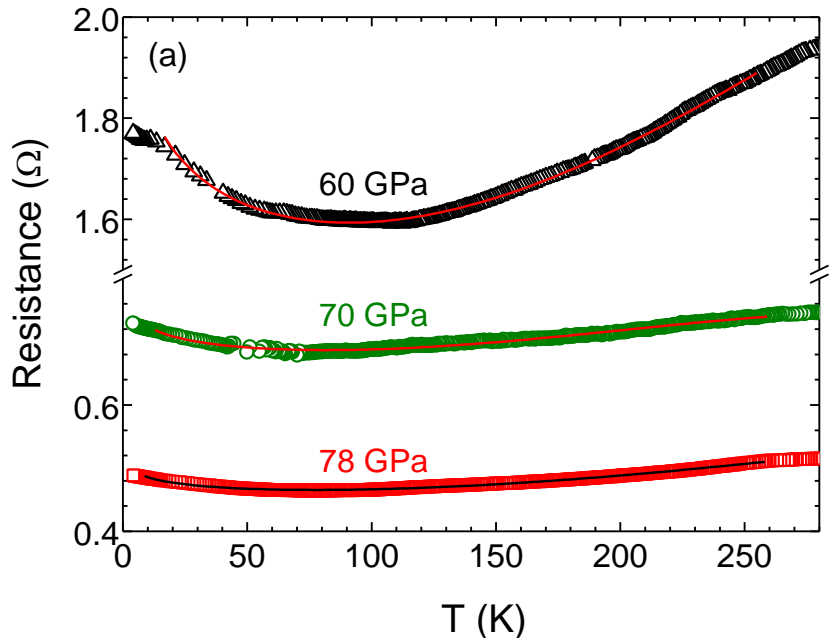



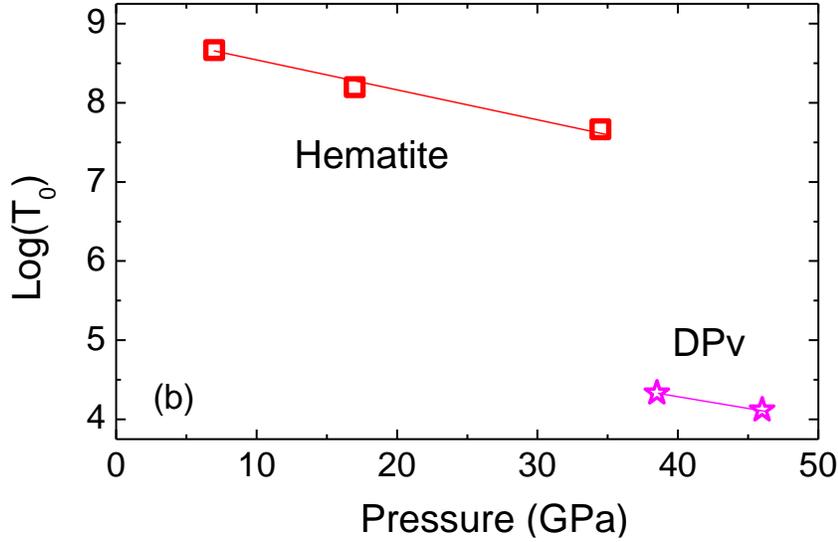

**SI Figure 6.** Temperature dependence of electrical resistance at various pressures in the metallic region (a) and pressure dependence of Mott temperature, $T_0$, for the hematite and insulating DPv phases (b). In the metallic region (a), above ~50 GPa, $R(T)$ exhibits a Fermi-liquid-like $R \sim T^2$ dependence with a minimum: at $T_{min} \approx 110$ - $150$ K in the DPv phase and at $T_{min} \sim 75$ K in the HP phase. This behavior could be a consequence of the Kondo effect – an indication of strong interaction between localized magnetic moments and the conduction electrons. We note the rather high value of $T_{min}$ for both phases. The solid lines represent results of the fit to $R(T) = R(0) + aT^2 + bT^5 - c\ln(T)$, where $R(0)$ is the sum of all contributions to the resistance at zero temperature.

In the insulating region (b), the temperature dependence of the resistance of the insulating DPv phase, below ~50 GPa, and the hematite phase is associated with a variable-range hopping mechanism ($\sigma = C \exp(T_0/T)^{1/4}$). We note ~ 4 order of magnitude difference of the Mott temperature between hematite and DPv phases.



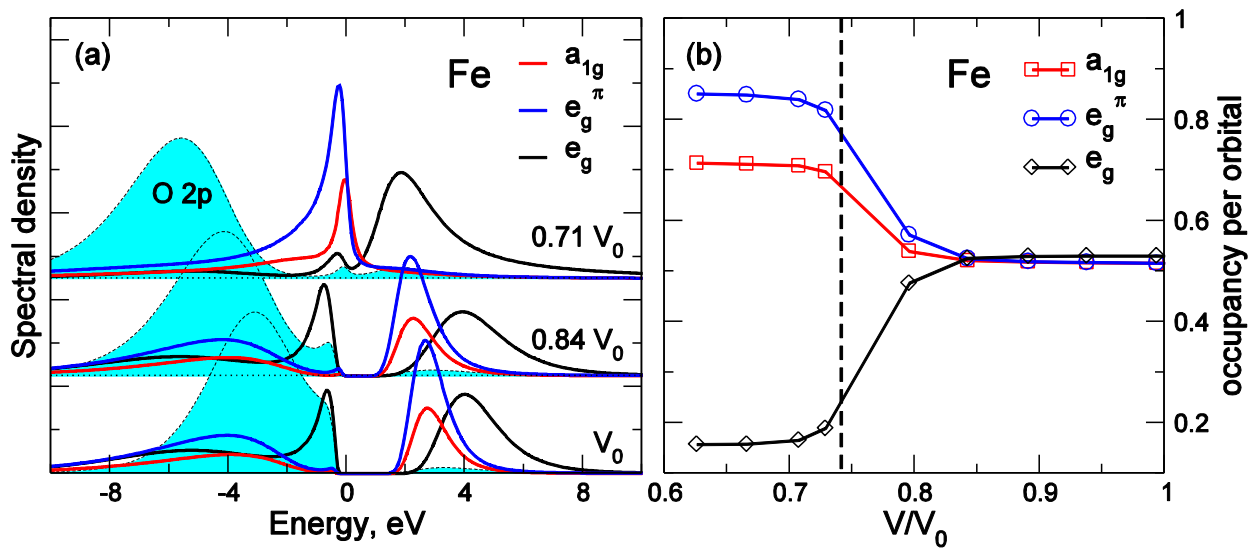

**SI Figure 7.** Evolution of the Fe *3d* and O *2p* spectral function (a) and the partial Fe $t_{2g}$ and $e_g$ occupations (b) of paramagnetic *R-3c* $Fe_2O_3$ calculated by DFT+DMFT at T=1160 K as a function of lattice volume. Fe $t_{2g}$ ($a_{1g}$ and $e_g^\pi$ orbitals) and $e_g$, and O *2p* orbital contributions are shown. The MIT associated with a HS-LS state transformation takes place at pressure ~ 72 GPa (at volume ~ 0.74 $V_0$).



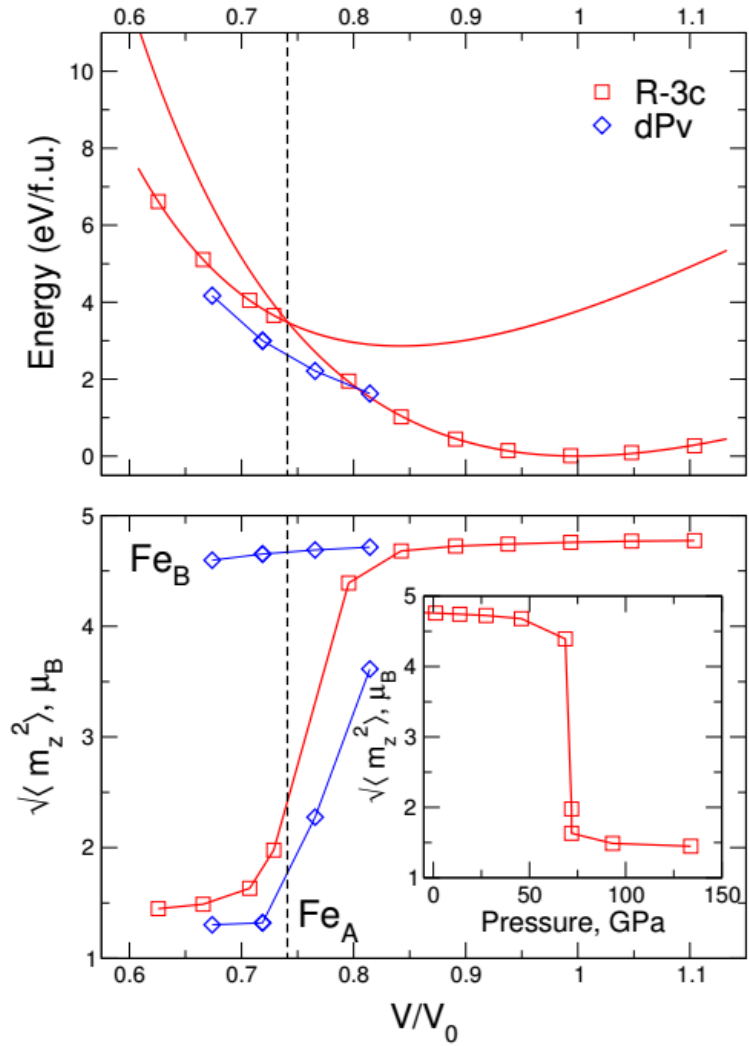

**SI Figure 8.** Total energy and local moment $\sqrt{\langle m_z^2 \rangle}$ of paramagnetic $Fe_2O_3$ calculated by DFT+DMFT at $T$=1160 K as a function of lattice volume. The lattice collapse associated with a HS-LS state transition is depicted by a vertical black dashed line. Our result for the DPv phase is depicted by a diamond.



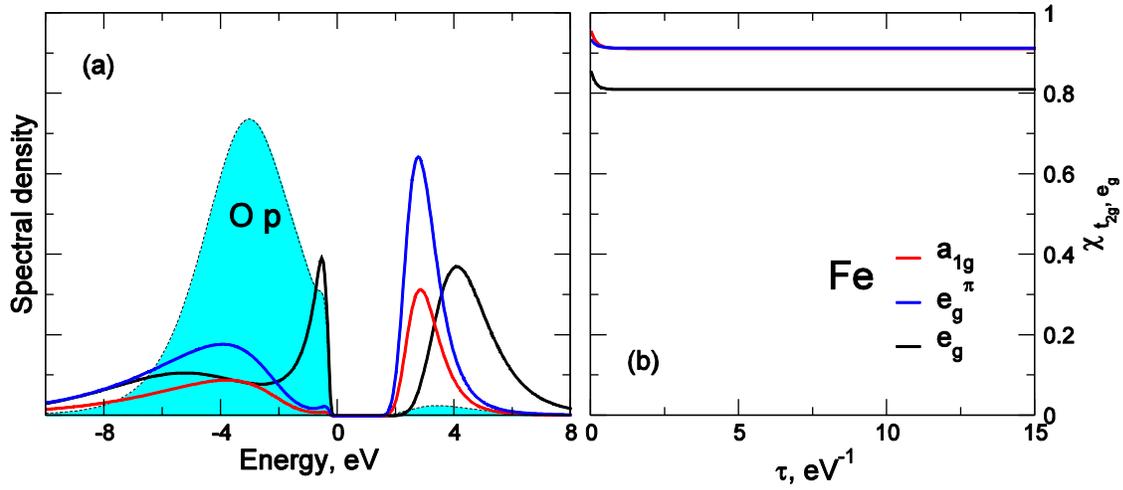

**SI Figure 9.** Fe *3d* and O *2p* spectral function (a) and spin-spin correlation function $\chi(\tau)=\langle m_z(\tau)m_z(0)\rangle$ (b) of paramagnetic corundum $Fe_2O_3$ as calculated by DFT+DMFT for the equilibrium volume $V_0$, at temperature $T$=390 K. Fe $t_{2g}$ ($a_{1g}$ and $e_g^\pi$ orbitals) and $e_g$, and O *2p* orbital contributions are shown. $\chi(\tau)$ is seen to be almost constant, independent of $\tau$, and close to the unit, implying a strong localization of the Fe *3d* electrons.

**SI Table 1:** Fe-O interatomic distances (Å) in the hematite (at 51.2 GPa), double perovskite (DPv) $P2_1/n$ (at 54.3 GPa)[62] and HP $Aba2$ (at 73.8 GPa) phases of $Fe_2O_3$ at ambient temperature

| Structure | Interatomic distances, Å | | | |
|---|---|---|---|---|
| | Fe−O1 | Fe−O2 | Fe−O3 | <Fe−O> |
| $R\bar{3}c$ | 1.832(2) | 1.989(3) | | 1.9105(25) |
| $P2_1/n$ | | | | |
| Site $B'$ | 1.85(2) x 2 | 1.82(2) x 2 | 1.92 (3) x 2 | 1.863(23) |
| Site $B''$ | 1.796(15) | 1.74(4) | 1.81(2) | 1.782(25 ) |



| Site $A$ | 1.87(3) | 1.910(19) | 1.89(3) | 2.087(26) |
|---|---|---|---|---|
| | 2.196(16) | 2.14(3) | 1.931(19) | |
| | 2.39(3) | 2.37(3) | | |
| $Aba2$ | 1.79(4) | 1.77(5) | | 1.817(40) |
| | 1.81(4) | 1.85(4) | | |
| | 1.83(3) | | | |
| | 1.85(4) | | | |

**Supplementary Table 2:** Structural data for the various phases used in the theoretical calculations

| Structure | a, Å | b, Å | c, Å | Additional | Fractional atomic positions | |
|---|---|---|---|---|---|---|
| | | | | | Fe | O |
| $R\bar{3}c$ | 5.0346 | 5.0346 | 13.7473 | | 0, 0, 0.35534 | 0.3056, 0, 0.25 |
| $P2_1/n$ | 4.588 | 4.945 | 6.679 | β=91.31° | Fe1: 0, 0, 0 | O1: 0.338, 0.185, 0.072 |
| | | | | | Fe2: 0, 0, 0.5 | O2: 0.325, 0.181, 0.430 |
| | | | | | Fe3: 0.5282, 0.0828, 0.7505 | O3: 0.852, 0.063, 0.262 |
| $Aba2$ | 6.524 | 4.702 | 4.603 | | 0.6094, 0.2095, -0.094 | O1: 0.355, 0.15, 0.049 |
| | | | | | | O2: 0.5, 0.5, 0.127 |

**DFT+DMFT calculations of the electronic and structural properties of the $R$-$3c$ phase of Fe$_2$O$_3$**

We calculate the electronic structure and phase stability of the corundum $R$-$3c$ phase of Fe$_2$O$_3$ using the fully charge self-consistent DFT+DMFT approach[62,62,62,62,62] implemented with plane-wave pseudopotentials[62,62,62,62]. To this end, we calculate the total energy and local moment of the Fe ions of the $R$-$3c$ phase as a function of lattice volume. The calculations are performed in a paramagnetic state at temperature $T = 1160$ K. We use the average Coulomb interaction $U = 6$ eV and Hund's exchange $J = $



0.86 eV for the Fe *3d* shell as was estimated previously[62]. The *U* and *J* values are assumed to remain constant upon variation of the lattice. Overall, our results for the electronic and lattice properties of the *R-3c* phase agree well with experimental data. We first discuss the spectral properties of paramagnetic $Fe_2O_3$. In Fig. S7 (a) we present the evolution of the spectral function of $Fe_2O_3$ calculated as a function of lattice volume. At ambient pressure, we obtain a Mott insulating solution with an energy gap of ~2.5 eV, in agreement with optical and photoemission experiments[62,62,62]. Upon compression, the energy gap gradually decreases, resulting in a Mott-Hubbard insulator-to-metal transition (MIT), which is associated with a high-spin (HS) to low-spin (LS) state transition**Error! Bookmark not defined.**. In fact, as shown in Fig. S7 (b), the MIT is accompanied by a remarkable redistribution of the Fe *3d* charge between the $t_{2g}$ and $e_g$ orbitals. Fe $t_{2g}$ orbital occupations are found to gradually increase upon compression. In particular, at a pressure above ~75 GPa, the $a_{1g}$ orbital occupancy is about 0.7, while the $e_g^\pi$ occupation ~0.85. On the other hand, the Fe $e_g$ orbitals are strongly depopulated (their occupation is below 0.2).

In Fig. S8 we show our results for the evolution of the total energy and local magnetic moment of paramagnetic $Fe_2O_3$ as a function of lattice volume. We fit the calculated total energy using the third-order Birch-Murnaghan equation of states separately for the low- and high-volume regions. Our results for the equilibrium lattice constant *a*=5.61 a.u. and bulk modulus $K_0$ ~187 GPa ($K_0$'=d$K$/d$T$ is fixed to 4.1) are in good quantitative agreement with the XRD data. At ambient pressure, the calculated local magnetic moment is ~ 4.76 $\mu_B$, implying a high-spin *S*=5/2 state of the $Fe^{3+}$ ions ($3d^5$ configuration with three electrons in the $t_{2g}$ and two in the $e_g$ orbitals). Our result for the spin-spin correlation function $\chi(\tau)=\langle m_z(\tau)m_z(0)\rangle$ calculated by DFT+DMFT for the equilibrium volume $V_0$ and *T*=390 K is seen to be almost constant, independent of $\tau$, and close to the unit (see Fig. S9). This implies that the Fe *3d* electrons are strongly localized to form fluctuating moments. Upon compression of the *R-3c* lattice to $V/V_0$ ~ 0.74, the total energy and local moment show a remarkable anomaly. In fact, the local moment is seen to retain its high-spin value down to about 72 GPa, while upon further compression, it exhibits a sharp HS-to-LS transition (see inset of Fig. S8), with a LS moment ~ 1.5 $\mu_B$ at pressure above ~ 90 GPa.

Our calculations reveal that the HS-LS transition in the *R-3c* structure of paramagnetic $Fe_2O_3$ is associated with a Mott-Hubbard MIT. Moreover, the MIT is accompanied by an isostructural collapse of the lattice volume by ~ 12%, implying a complex interplay between electronic and lattice degrees of freedom. The structural change takes place upon compression above ~ 72 GPa. In addition, we find that the bulk modulus in the HS phase ($K_0$ ~187 GPa) is considerably smaller than that in the LS phase (245 GPa), resulting in a remarkable decrease of the compressibility at the phase transition.

Furthermore, our theoretical results for the *R-3c* phase of $Fe_2O_3$ show that the MIT occurs at a remarkably high pressure value of ~72 GPa. It is considerably higher than the structural transformation into the DPv phase found experimentally (~50 GPa). We also calculate the total energy for the DPv phase of paramagnetic $Fe_2O_3$ using the crystal structure parameters obtained at ~54 GPa. Our results reveal that the DPv phase is energetically favorable in comparison to the corundum *R-3c*, i.e., the DPv phase is thermodynamically stable under pressure. In addition, as discussed in our paper, we obtain that the DPv phase is a site-selective Mott insulator, in which the *3d* electrons on only half of the Fe sites (octahedral *B* sites) of DPv $Fe_2O_3$ are metallic, while the others (*A* sites) remain insulating. Our theoretical results for the phase stability $Fe_2O_3$ thereby confirm a structural transition from the corundum *R-3c* to the DPv structure of $Fe_2O_3$, in agreement with our experimental data. Overall, our results for the electronic structure, equilibrium lattice constant, and structural phase stability of paramagnetic $Fe_2O_3$ agree remarkably well with experimental data.

**Evolution of the transport properties across the insulator-to-metal transition**



Evolution of the transport properties across the transition can be interpreted similar to Machavariani *et al.*[62] by assuming that the sample is a mixture of two phases, metallic and insulating, with different transport characteristics. The overall conductivity σ is determined by a relative volume of both phases and by the shape and distribution of the clusters of each phase. We can estimate the relative volume of each phase from the room temperature resistivity measurements, assuming roughly that the clusters are spherical and that the conductivities of the two phases, $\sigma_1$ and $\sigma_2$, are not changed across the transition. In the framework of the symmetrical effective medium theory of Bruggeman[62], the relative volumes $V_1$ and $V_2 = 1-V_1$ are given by

$$V_1 \frac{\sigma_1 - \sigma}{\sigma_1 + 2\sigma} + (1 - V_1) \frac{\sigma_2 - \sigma}{\sigma_2 + 2\sigma} = 0 \quad . \quad (1)$$

Fig. 2c shows the relative volume of metallic phase (the *HP* abundance) as deduced from Eq. (1) for decompression and recompression cycles. The $\sigma_1$ and $\sigma_2$ values are chosen as the estimated conductivities just before and after the transition, where the phases exist alone (at ~40.6 GPa and ~70.4 GPa for decompression and at 47 GPa and 76.6 GPa for recompression). The calculations are made on the assumption that the geometrical coefficient $B$ in the relation $\sigma = B/R$ does not change under pressure ($B$ depends on the thickness of the sample, the distance between the contacts and the width of the current flow).